\documentclass[10pt,letterpaper]{article}
\usepackage[top=0.75in,left=0.75in,footskip=0.75in]{geometry}
\usepackage{amsmath,amssymb}

\usepackage{changepage}

\usepackage[utf8x]{inputenc}

\usepackage{textcomp,marvosym}

\usepackage{cite}

\usepackage{nameref,hyperref}

\usepackage[right]{lineno}

\usepackage{microtype}
\DisableLigatures[f]{encoding = *, family = * }

\usepackage[table]{xcolor}

\usepackage{array}

\newcolumntype{+}{!{\vrule width 2pt}}

\newlength\savedwidth

\raggedright
\usepackage[aboveskip=1pt,labelfont=bf,labelsep=period,justification=raggedright,singlelinecheck=off]{caption}

\makeatletter
\renewcommand{\@biblabel}[1]{\quad#1.}
\makeatother

\usepackage{lastpage,fancyhdr,graphicx}
\usepackage{epstopdf}
\fancyheadoffset[L]{2.25in}
\fancyfootoffset[L]{2.25in}
\begin{document}
\vspace*{0.2in}

\begin{flushleft}
{\Large
\textbf\newline{On optical transmittance of ultra diluted gas}
}
\newline
\\
Jakub M. Ratajczak
\\
\bigskip
Centre of New Technologies, University of Warsaw, Poland
\\
\bigskip

% Use the asterisk to denote corresponding authorship and provide email address in note below.
j.ratajczak@cent.uw.edu.pl

\end{flushleft}

\section*{Abstract}
The paper proposes a model of optical transmittance of ultra diluted gas taking into account gas particles non-locality, the quantum effect of wave function spreading derived from solving the Schrödinger equation for a free particle. A significant increase in the transmittance of such gas is envisaged as compared to the classical predictions. Some quantitative and qualitative consequences of the model are indicated and falsifying experiments are proposed. The classic Beer-Lambert law equation within range of its applicability is derived from the model. Remarks to some astrophysical phenomena and possible interpretations of Quantum Mechanics are made. An experiment consistent with the predictions of this model is referenced.

\section*{Introduction}
The appropriate models of light-gas interaction are crucial for many fields of science like astrophysics, atmospheric science, chemistry or cosmology. There are many theories describing transmittance, absorbance, scattering and other phenomena of electromagnetic waves passing by clouds of particles. One of the oldest is the Beer-Lambert exponential transmission law \cite{Bouguer1729} \cite{A.D.McNaught1997} describing attenuation of monochromatic light by the homogeneous, not very dense medium. It is still used for many applications, mostly for quantitative spectroscopy \cite{Bernath2016}. Modern models are far more advanced. They are in a form of transfer equations \cite{DavisA.2005}, they may be applicable for non-heterogeneous media \cite{Kostinski2001}. Some of them are derived directly from Maxwell equations \cite{Mishchenko2003}, some may cover multiple scattering etc. To our best knowledge, this sort of theories are mostly derived from some kinds of an ideal gas model. Gas particles are treated as small localized entities of finite size. 

In this paper we propose a model of an ultra diluted gas transmittance measurement. We assume that molecules of a gas, where mean free time is long enough, should follow the Schrödinger equation for a free particle \cite{Shankar2011} when in non-relativistic limit. We refer to particle free time as a time between \textit{any} interaction with a particle: either with a photon or with another particle. This way we can discard the potential term in the Schrödinger equation and apply the free particle equation. The well known solution of this equation renders to a serious particle’s wave function spatial spreading over time. Our intention is to investigate how the measured transmittance is influenced by the relation of spreading and the size of a light detector. Besides, an increasing number of experiments \cite{Handsteiner2017}\cite{Rauch2018} convince us that Quantum Mechanics is not a local realistic theory \cite{Einstein1935} \cite{Musser2016} so we assume wave function non-locality.

The proposed model displays an interesting property that measured transmittance grows with particles' mean free time, despite gas density kept constant. It is also shown that measured transparency of a cloud depends on a detector size and on a background radiation intensity a gas cloud is exposed to.

A variety of astrophysical phenomena may relate to this properties. A number of falsifying experiments, including laboratory table top experiments, are proposed. Also, the limitations of the applicability of some classic laws are pointed out.

In the next section we outline necessary assumptions. In the Section 3 the model of transmittance of a cloud of spreaded particles is proposed. There is the derivation of the classic Beer-Lambert law presented in the fourth section. Section 5 includes a basic analysis, where a stable ``smeared'' gas may occur. The final sections contain suggestions for falsifying experiments, references to some known phenomena and the summary. The Appendix includes a less numerically demanding form of the smeared gas transmittance formula.

\section*{Assumptions}
\subsection*{Gas particles model}
In the proposed model an internal structure of gas particles is discarded. We treat each of particle as an individual, independent Gaussian wave packet of a given mass. 

The geometrical cross-section of a photon-particle scattering is used later in the paper as a simplification of internal details of any process that influences photon propagation i.e., absorption or scattering. It is justified as the final transmittance model refers to a cloud of particles. In the diluted ensemble of particles, in thermal equilibrium, all individual shapes, rotations, scattering angles, energy states etc. may be averaged e.g., with a geometrical cross-section. A probability of each single scattering event is assumed to be identical. The certain geometrical cross-section may be also understood as a notion of constant Einstein coefficients \cite{Bernath2016}. 

Assumption of weak light and thermal equilibrium allows to omit phenomena like stimulated emission, energy state inversion, etc. They may be considered later on as higher order corrections, however. It is also possible to extend later the proposed transmittance model to more sophisticated cases of non-heterogeneous scattering media.

It is unfortunate that standard deviation and total cross-section are designated with the same symbol ``$\sigma$'' in the literature. We will be using ``$\sigma$'' to designate the total cross-section and ``$stdev$'' for standard deviation.

\subsection*{Non-locality}
The key assumption of the proposed model is that wave functions of gas molecules are non-local. We are convinced by a growing number of experiments confirming Quantum Mechanics non-locality e.g. quantum entanglement \cite{Einstein1935}\cite{Handsteiner2017} \cite{Rauch2018}, Bose-Einstein condensate\cite{Anderson2008}, helium Young’s type double-slit experiment \cite{Carnal1991}, etc. One of inspirations for this paper was the old work by Mott \cite{Mott1995} \cite{Figari2013}, an example how a non-local wave function may manifest as a localized particle. Moreover, recently a number of experiments confirm the occurrence of quantum effects in galactic scales e.g. \cite{Handsteiner2017}.
The paper follows interpretation of $|\Psi|^2$ as a pure probability not involving a corpuscle existence. 

\subsection*{Non-relativistic limit}
It was decided to apply the non-relativistic Schrödinger equation because the model is developed for clouds of relatively low energy particles. Gravity in the model is treated purely as a space-time geometry.

\subsection*{Gas particle spreading}
The well known one-dimensional wave function for a stable, non-relativistic gas particle \textit{A} reads as \cite{Shankar2011}:
\begin{equation}
\Psi_A(x',t=0)=e^{ip_0x'/\hbar}\frac{e^{-x'^2/2\Delta^2}}{(\pi\Delta^2)^{1/4}}~, \label{Psi1D}
\end{equation}
where $x'$ denotes position, $t$ time and $\hbar$ reduced Planck’s constant. The mean position $\langle X \rangle=0$ and its uncertainty $\Delta X=\Delta/2^{1/2}$ as well as the mean momentum $p_0$ and its uncertainty $\hbar/{2^{1/2}\Delta}$ corresponds to this packet. If it is not a subject of any reaction for some period of time $t$ the free motion evolution operator $U(x, t;x')$ may be applied:
\begin{equation}
U(x,t;x')= \left( \frac{m}{2\pi\hbar it} \right)^{1/2} exp\left( \frac{im(x-x')^2}{2\hbar t} \right)~, \label{U}
\end{equation}
where $m$ denotes particle mass. 

Applying the evolution operator to Eq.~(\ref{Psi1D}) and calculating probability density $P_A(x,t)=|\Psi_A(x,t)|^2$ we have a Gaussian:
\begin{equation}
P_A(x,t)=\left( stdev_A(t)\sqrt{2\pi} \right)^{-1/2} exp \left( \frac{-(x-p_0t/m)^2}{2stdev_A(t)^2} \right) \label{ProbDensity}
\end{equation}
with standard deviation $stdev_A(t,m,\Delta)$, where $m$ and $\Delta$ are assumed constant:
\begin{equation}
stdev_A(t)=\sqrt{\frac12 \left( \Delta^2+\frac{\hbar^2t^2}{m^2\Delta^2} \right)}\approx\frac{\hbar t}{\sqrt2m\Delta}~. \label{Stdev}
\end{equation}

A generalization for three dimensions assumes an independent increase of the spreading for each dimension. Moreover, we will be assuming further on, without prejudice to the generality of the considerations, that the initial measurement uncertainty in each direction $i=1,2,3$ is identical and equal to $\Delta$: $\Delta_i=\Delta$

\subsection*{Initial position uncertainty value}
In the actual physical setup, a particle position $\Delta$ measurement always has some specific uncertainty. We will stick to the assumption that $\Delta$ is constant, at least for a fair amount of time. For the qualitative considerations, as presented in this paper, it is not important what the exact figure is unless it's constant. However, it is extremely important for the quantitative analyses: performing calculations and planning experiments. 

Referencing e.g. \cite{Eichmann1993} we will assume that yet a single collision of an atom with a photon may lead to the wave function measurement.

\subsection*{Detector}
It's assumed that the light detector used for measuring transmittance is a kind of apparatus of a finite active area dimensions e.g., an eye, a telescope, a chip etc. Although a real device never has $100\%$ efficiency we will assume so. This is because a real efficiency of a detector can be usually easily determined and light measurements may be normalized as it would have $100\%$ efficiency.

\section*{Transmittance model}
\subsection*{Markov chain model}
With the assumptions from the previous section we can define dilute gas cloud transmittance $TR$ as follows. It is the probability that photon $\gamma$ coming from source $S$ \textit{that would have been detected by the detector $D$ in absence of a cloud} passes the entire $N$-element gas cloud without any collision $A_n{-}\gamma$ and is detected by detector $D$. Collisions with individual particles $A_n$ are independent, so we may consider this as a Markov chain and record such probability as a product of probabilities $(1-P_{{A_n}\gamma})$:
\begin{equation}
TR = \prod_{n=1}^N (1-P_{{A_n} \gamma})~, \label{TREquation}
\end{equation}
where $P_{{A_n} \gamma}$ is a probability of scattering a photon by $n$-th molecule of gas. Due to the transmittance definition above a scattering event referred by $P_{{A_n}\gamma}$ needs to occur in a volume, where a photon passing from source $S$ to detector $D$ is likely to be found. For a macroscopic setup it will be a set of classical trajectories.

\subsection*{An exemplary macroscopic setup}
Let's consider such an exemplary macroscopic setup similar to a typical astronomical measurement. A cloud of particles $A_n$ is at long distance $l_1$ from source $S$ of $\lambda$-wavelength photons $\gamma$. A detector $D$ is $l_2$ away from the cloud. They are all co-linear in regards to a photon passing by, so even if they move (relative to each other) their position is co-linear at each moment of a photon i) sent from the source, ii) passing by the particle, iii) reaching the detector. Thanks to non-relativistic limit we will consider particles' positions constant in regards to the axis $S{-}D$ (at least during a photon is passing by a particle). Both distances $l_1$ and $l_2$ are much greater than i) any expected standard deviation of a particle wave packet spread: $\forall t~l_{1,2}\gg stdev_A(t)$ and ii) cross-section $\sigma_{A\gamma}$ ``diameter'': $l_{1,2}\gg \sqrt{\sigma_{A\gamma}}$, iii) diameter of the cloud. This is to ensure that neither source nor detector fails into a region where particle may be found. Let's assume the detector has cross-section $\sigma_{D\gamma}$ towards $S$ and $100\%$ efficiency. The detector is of macroscopic size: $\sigma_{D\gamma} \gg \sigma_{A\gamma}(\lambda)$. For simplicity we will not consider a possible redshift assuming $\lambda=const$ all the way but one may easily extend the model considering distinct cross sections: $\sigma_{A\gamma}(\lambda_1)$ and $\sigma_{D\gamma}(\lambda_2)$.

The point size of the source (instead of some non zero cross-section towards the detector) may be justified by a possibility of re-scaling the model if necessary. It may require adjusting power of the source, the particle cross-section or the particle spreading speed in a plane perpendicular to the axis $S{-}D$ but one may easily extend the model. Also, the model may be extended to a full chromatic source. 

All this assumptions may be encoded as a single coefficient $G$ depending on $l_1, l_2, \sigma_{A\gamma}, \sigma_{D\gamma}, \lambda$. Important note is that this coefficient is time independent for each non-spread particle $A_n$. We will demand it to be (dimensionless) average probability of non detecting a photon sent from source due to single particle scattering. One may choose either i) a spherically symmetric source and use the detector cross-section considering conditional probability of a photon reaching a not obscured detector or ii) consider only photons that would reach detector anyway (a laser like source or a ``pencil'' beam). For simplicity we will follow the latter approach. For quantitative considerations it may require adjusting power of the source. With the latter approach we can disregard $l_2$ and $\sigma_{D\gamma}$.

For the above ``astronomical'' setup the coefficient $G$ reads as:
\begin{equation}
G(l_1, \sigma_{A\gamma}, \lambda)\approx \frac{\sigma_{A\gamma(\lambda)}}{4\pi l_1^2}~,
\end{equation}
where denominator is area of a sphere with radius $l_1$ and center in $S$.

The coefficient $G$ may be understood as a constant encoding the actual \textit{setup geometry}.

\subsection*{Smeared gas transmittance model}
In this section we will examine qualitative properties of the transmittance rule for a very diluted gas consisting of rarely colliding particles.

\subsubsection*{Smeared gas}

Let us consider a gas cloud made up of rarely, identical particles $A_n$, e.g. in the region of single digit particles per $\textrm{cm}^3$. Very few collisions $A{-}A$ will be taking place. Let us assume that this gas is located far from any sources of radiation, i.e. very few $A{-}\gamma$ scattering events will be taking place. Subsequently we will assume, introducing a simplification, that collisions $A{-}A$ and $A{-}\gamma$ are the only collisions that gas particles are subjected to. Later, heterogeneous gases may also be considered. However, for the qualitative considerations differentiating between various types of collisions is not relevant.

Let us note that for any given setup we can determine a mean time $\bar{t}$ between successive collisions of particles of the gas either \textit{with themselves} or \textit{with photons passing by}. This is the \textit{extended} mean free time. Let us not enter here in a discussion on what happens after the collision/measurement or what type of collision/measurement causes a collapse of the wave function of particles $A_n$. It is just important that particles have some significant free time $\bar{t}$ to apply Eq.~(\ref{U}). This way the expected value of standard deviation $stdev_{A_n}(\bar{t})$ of particles $A_n$ spreading will be proportional to the expected value of extended mean free time $\bar{t}$, with the other parameters being constant. We will call this type of gas the \textit{smeared gas}.

Let us consider what the transmittance of such gas on line $S$-$D$ could be? Whether, and how it could be dependent on the spreading of the wave functions of the gas particles?

\subsubsection*{Detectability tunnel}

As discussed above we can see that for the detector $D$ to ``detect'' an occurrence of a collision (i.e. photon not reaching the detector), it must occur in a certain, precisely defined volume. For the macroscopic setup this volume is a set of classical paths of a photon from $S$ to $D$ in a shape of a pyramid with source $S$ as its apex and the detector, as its base. We will call this volume the \textit{detectability tunnel}. 

Due to easier analytic forms of integrals used in further calculations let us assume that the detector’s shape is a square with a side designated as $d_T=\sqrt{\sigma_{D\gamma}}$. Let us assume that the detector is much larger than atomic scales and much smaller than $l_1$ and $l_2$: $1[\textrm{Å}] \ll d_T \ll l_{1,2}$. These assumptions do not affect the key conclusions.

We will adopt one more simplification that will allow fully analytic approach. In spite of the note on source $S$ being a point, we will assume that the tunnel has a constant cross-section along its entire length: cross-section equal to the square cross-section of the detector. As it will turn out this simplification does not change the essence of the argument. Due to the differences of scales ($d_T \ll l_{1,2}$, $d_T \ll \sigma_A(t)$) they will have to be taken into account only in the case of very accurate calculations. This assumption also defends itself one other way. In the enlarged tunnel particle's wave function $\Psi_A$ will be occurring with a higher probability – as a cubic tunnel take up more space than the pyramid inscribed therein. Therefore, if it is possible to prove that transmittance increases taking into account the probabilities in the larger cubic tunnel, then transmittance will increase even more in the tunnel inscribed therein – as the probabilities of any obstacles occurring will be lower there.

We may treat light paths classically because the tunnel is of a macroscopic scale comparing to light wavelengths. Bringing a path integrals formalism: amplitudes of photon paths out of the tunnel cancel each other. Or refer this simplification to  simplification of the wave optics by the geometric optics \cite{Hecht2015}. Besides, common sense observations shows that macroscopic objects out of a classic light path do not overlap distant light sources.

\subsubsection*{A single particle transmittance}

\begin{figure}
\includegraphics[width=3.5in]{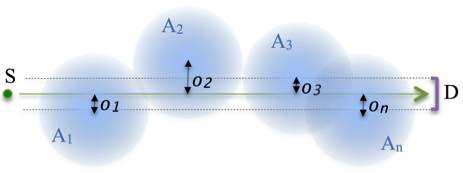}
\caption{Positions of probability distributions $|\Psi_{A_n}(x,t)|^2$ of gas particles $A_n$ in relation to axis $S{-}D$. The detectability tunnel is marked with dotted lines. In a macroscopic setup only events occurring within the tunnel may prevent a photon from reaching detector $D$.} \label{fig:fig_position_of_gas_particles}
\end{figure}

FIG.~(\ref{fig:fig_position_of_gas_particles}) shows a few sample gas particles probability distributions (gaussians) positions along with a range of the detectability tunnel – between dotted lines. Assuming the macroscopic setup the tunnel is the only volume where a \textit{detectable} scattering event may occur. Therefore, one can see that detector $D$ detects collision $A_n-\gamma$ with probability $P_{A_n\gamma}$ that is proportional to the probability density constrained by tunnel $T$:
\begin{equation}
P_{A_n\gamma}(t,o_n) = G() \int_T |\Psi_{A_n}(\mathbf{r},t)|^2dr~, \label{P_Ag}
\end{equation}
where $t$ is the particle's proper time running from the last measurement. It is clear that the above integral is equal to 1 (Dirac's delta) in case of the non spread particle $A_n$.

We assume the $\sigma_{A_n\gamma}$ is constant during spreading: $\forall{t}~\sigma_{A_n\gamma}(t)=\sigma_{A_n\gamma}(0)$. It makes $G()$ coefficient constant. This is assuming particle’s Einstein coefficients unchanged during spreading. Getting ahead of the argument, it should be mentioned that a sufficient precondition for the validity of this paper’s thesis is for $\sigma_{A_n\gamma}(t)$ not to rise too fast over time.

Probability $P_{A_n\gamma}(t,o_n)$ is spreading time $t$ dependent due to the fact that the probability of finding a particle in tunnel $T$ depends on spreading time.

Let us note that Eq.~(\ref{P_Ag}) is directly in line with probabilistic interpretation of Quantum Mechanics. It takes into account the stipulation related to the meaning of the wave function squared as the probability of particle’s position in space.
The detectability tunnel does not have to run through the center of particle spreading area. Its offset will depend on the expected value of particle position in relation to the $S{-}D$ axis. However, it can be noted that probability $P_{A_n\gamma}(t,o_n)$ will be the highest for the tunnel running centrally i.e., $o_n=0$ – due to the fact that it runs through the densest part of particle probability density.

Finding the analytic form of probability $P_{A_n\gamma}(t,o_n)$ seems challenging. Especially when wishing to take into account various shapes of detectors or the exact conical shape of the tunnel. Instead, we will look for the upper limit $P_{A_n\gamma}^{max}(t,o_n)$ of probability $P_{A_n\gamma}(t,o_n)$:
\begin{equation}
P^{max}_{A_n\gamma}(t,o_n): \forall{t,o_n} ~ P^{max}_{A_n\gamma}(t,o_n) \geq P_{A_n\gamma}(t,o_n)~. \label{Pmax_to}
\end{equation}
Note, that $d_T$, $m$ and $\Delta$ are all constant.

Let us take the Cartesian coordinates with axis Z running parallel to axis $S{-}D$. We may write $P^{max}_{A_n\gamma}(t,o_n)$ for 3 dimensions:

%\begin{widetext}
\begin{equation}
P^{max}_{A_n\gamma}(t,o_n)=G()~\int_{-\infty}^{\infty} \int_{o_n-r_T}^{o_n+r_T} \int_{o_n-r_T}^{o_n+r_T} |\Psi_{A_n}(x,y,z,t)|^2 dx~dy~dz \label{PmaxExplicit}
\end{equation}
%\end{widetext}
where $r_T=d_T/2$ is the square detector ``radius''.

We have assumed the limits of $dz$ integration as $\pm \infty$ due to the fact that while conducting considerations in macroscopic scales we may assume that distance $l_1+l_2$ between the source and the detector is much larger than the spreading of particle $A_n$: $l_1+l_2 \gg stdev_{A_n}(t)$ for any $t$. First of all, this will simplify considerations and further formulas. Secondly, the far ends of the normal distribution are asymptotically driving to 0 and make a negligible contribution. And thirdly, even if they make a contribution, it will only increase the probability of an occurrence of a collision reducing the transmittance which is permitted when estimating the upper limit of $P_{A_n\gamma}(t,o_n)$. We allowed ourselves to adopt the Cartesian coordinates, disregarding potential space-time curvature, as we may assume that we are conducting considerations in the area where space-time is sufficiently flat. If necessary, the same technique may be used for a non-flat space-time.

Disregarding the momentum of particles $A_n$ (i.e., inserting $p_0=0$ as $v_{\gamma} \gg v_A$ and using Eq.~(\ref{ProbDensity})) we can provide a well known, three-dimensional, independent in each dimension, probability distribution for a free particle $A_n$:

%\begin{widetext}
\begin{equation}
P_{A_n}(x,y,z,t)=|\Psi_{A_n}(x,y,z,t)|^2= \left( \frac{1}{\sqrt{2\pi}stdev_A(t)}\right)^3exp \left( \frac{-x^2-y^2-z^2}{2stdev_A(t)^2}\right)~, \label{P_A_n}
\end{equation}

where, in accordance with the convention, we have designated standard deviation as $stdev_{A_n}(t)$ according to Eq.~(\ref{Stdev}).

Applying Eq.~(\ref{P_A_n}) to Eq.~(\ref{PmaxExplicit}) analytic form of $P^{max}_{A_n\gamma}(t,o_n)$ for a square detector and constant tunnel cross-section reads as:
\begin{equation}
P^{max}_{A_n\gamma}(t,o_n)=G()~\frac{1}{4} \left[ erf \left( \frac{o_n-r_T}{\sqrt2 stdev_{A_n}(t)} \right) - erf \left( \frac{o_n+r_T}{\sqrt2 stdev_{A_n}(t)} \right) \right]^2~, \label{PmaxAnalytic}
\end{equation}
where $erf()$ denotes the Gauss error function. 
%\end{widetext}

Eq.~(\ref{PmaxAnalytic}) is plot in FIG.~(\ref{fig:uppper_limit_Pmax_to}) assuming that parameters $\Delta$, $m$, $r_T$ and Einstein coefficients ($G(t)=const$) are all constant. No matter where particle is located, thanks to $stdev_{A_n}(t)$ rising over particle free time the probability of a photon being obscured by particle $A_n$ is decreasing asymptotically to zero:

\begin{figure}
\includegraphics[width=3in]{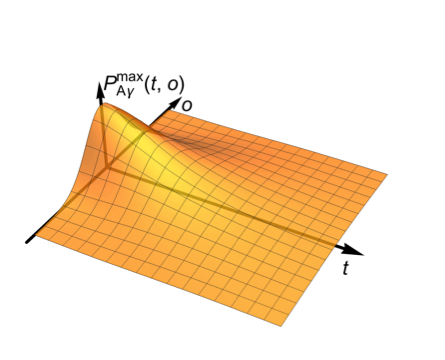}
\caption{Upper limit of probability $P^{max}_{A_n\gamma}(t,o_n)$ of an occurrence of collision $A_n{-}\gamma$ in the function of particle free time $t$, and distance $o_n$ from the source-detector axis, arbitrary units, linear axis.} \label{fig:uppper_limit_Pmax_to}
\end{figure}

\begin{equation}
\forall{o_n} \lim_{t\to\infty} P^{max}_{A_n\gamma}(t,o_n)=0~. \label{LimesPmax}
\end{equation}

Together with Eq.~(\ref{Pmax_to}) and Eq.~(\ref{LimesPmax}) it yields to:
\begin{equation}
\forall{o_n} \lim_{t\to\infty} P_{A_n\gamma}(t,o_n)=0~. \label{LimesP}
\end{equation}

This result demonstrates that over (particle free) time it becomes more and more difficult to find particle $A_n$ within specific, fixed volume. A detector of finite size is able to locate events in some finite volume only. This way the effective total cross-section of a spreading particle is actually decreasing over (particle free) time from the point of view of a finite detector.

Due to physical setup constraints there is always some upper limit for $t$, at most the age of the Universe. For a gas particle it is the \textit{extended} mean free time. Note, that the detector $D$ should be small enough for relation $r_T < stdev_{A_n}(t)$ to be held for considered particle free time $t$. If the detector is too big (causing $erf()$ function argument to be greater than 2 for any $t$ considered) there will be no effect of decreasing effective cross section.

In general the tunnel does not have to be straight. It will be straight only based on the assumption that particle $A_n$ does not move orthogonal to axis $S{-}D$ and disregarding such effects as, for example, space-time curvature. However, the tunnel’s shape is not important from the point of view of the qualitative discussion. It is only important that the tunnel does not include within itself the entire area where particle $A_n$ may be found: $r_T < stdev_{A_n}(t)$. Irrespective of the tunnel’s shape – with its fixed volume – Eq.~(\ref{Pmax_to}) is held.

Due to the fact that $l_1+l_2 \gg stdev_{A_n}(t) \gg d_T$, the detector's shape (tunnel cross-section) does not have a significant impact on the value of the probabilities calculated above. Certain approximations can be made in the calculations, if required. For example, if a detector with a circular cross-section with diameter $d_T$ were used, then the probability could, with a good approximation, be decreased by $\pi/4$.

The range of variable $t$ (i.e. a particle's time) requires a comment. One can assume values from $0$ to the age of the Universe. Therefore, we will never reach infinity in Eq.~(\ref{LimesPmax}). However, inserting specific values into the calculations, especially for light molecules like hydrogen and helium, it turns out that even already for small $t$ values $P_{A_n\gamma}$ reaches such low values that the effective total cross-section practically drops to zero. On the other hand, the upper limit of $t$ allows us to safely write such inequalities as: $l_1+l_2 \gg stdev_{A_n}(t)$.

We have shown that from the point of view of an observer equipped with a finite detector the probability of interaction between a probe particle (e.g. a photon $\gamma$) and particle $A_n$ will weaken as a result of particle wave function spreading.

Let us note that above considerations are valid not just for photons and gas molecules. They can be generalized with respect to other types of collisions of various physical objects (including macroscopic ones) represented by wave functions.

\subsubsection*{Transmittance of smeared gas}

Let's define $P_{G\gamma}$ as a probability of scattering photon $\gamma$ by any gas particle $A_n$. This is photon not passing through the cloud. It is a complement of the transmittance Eq.~(\ref{TREquation}):
\begin{equation}
P_{G\gamma} = 1-TR = 1 - \prod_{n=1}^N (1-P_{{A_n}\gamma})~. \label{P_G}
\end{equation}

In this non-local model we need to take into account the impact of all particles of a gas cloud, and not only those particles whose expected value of position is located inside tunnel $T$. This is due to the fact that if $stdev_{A_n}(\bar{t}) \gg d_T$, and even if $stdev_{A_n}(\bar{t}) > d_T$, then the particles positioned \textit{near} the tunnel may have a significant impact on the probability density in the tunnel. Therefore, we should think of \textit{a gas cloud as a collection of gas particles that are positioned close enough to the tunnel, so that one cannot disregard the impact of the probability of finding them in the tunnel}. In general one needs to take into account all particles in the Universe. However, the distribution of many of them will be similar to the Dirac delta or they will be positioned too far ($o_n \gg stdev_{A_n}(t)$), to make a significant contribution to the integral of the probability density in the tunnel $T$: $\int_T|\Psi_{A_n}|^2 \cong 0$.

Let us try to analyze extended mean free time dependency of the probability of any collision $A_n{-}\gamma$ in detectability tunnel $T$ running through the (homogeneous) smeared gas cloud. We will designate this probability as $P_{G\gamma} (\bar{t})$. It is related to transmittance $TR$ of the cloud with relationship $TR=1-P_{G\gamma} (\bar{t})$. Note that $P_{G\gamma} (\bar{t})$ is not absorbance. 

Now, let's find upper limit $P_{G\gamma}^{max} (\bar{t})$ of probability $P_{G\gamma} (\bar{t})$:
\begin{equation}
P_{G\gamma}^{max} (\bar{t}): \forall{\bar{t}}~P_{G\gamma}^{max} (\bar{t}) \ge P_{G\gamma} (\bar{t})~. \label{P_G_req}
\end{equation}

Taking Eq.~(\ref{P_G}) let's consider the following form of $P_{G\gamma}^{max} (\bar{t})$:
\begin{equation}
P_{G\gamma}^{max} (\bar{t}) = 1- \prod_{n=1}^N \left(1-P_{A_n \gamma}^{max} (\bar{t}, o_n) \right)~. \label{P_G_max}
\end{equation}

With Eq.~(\ref{Pmax_to}) we may show:
\begin{equation}
\forall{\bar{t}, o_n}~P_{A \gamma}^{max} (\bar{t}, o_n) \ge P_{A_n \gamma} (\bar{t}, o_n)~.
\end{equation}
Therefore Eq.~(\ref{P_G_max}) fulfills the requirement of Eq.~(\ref{P_G_req})

Now, using Eq.~(\ref{P_G_max}) and Eq.~(\ref{PmaxAnalytic}), we may provide the analytic formula for the \textit{minimum} value of transmittance $TR(\bar{t})$ of smeared gas cloud for the square detector with side $d_T=2r_T$:

%\begin{widetext}
\begin{equation}
TR(\bar{t}) \ge \prod_{n=1}^N \left( 1-\frac{G()}{4} \left[ erf \left( \frac{o_n-r_T}{\sqrt2 stdev_{A_n}(\bar{t})} \right) - erf \left( \frac{o_n+r_T}{\sqrt2 stdev_{A_n}(\bar{t})} \right) \right]^2 \right)~. \label{TRUpperLimit}
\end{equation}
%\end{widetext}

\begin{figure}
\includegraphics[width=3.5in]{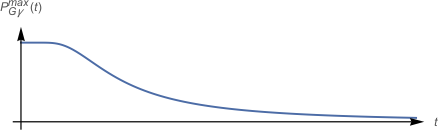}
\caption{Upper limit of probability $P_{G\gamma}^{max} (\bar{t})$ that a finite size detector is able to notify photon scattering event in a smeared gas cloud as a function of gas particles extended mean free time, arbitrary units, linear axis.} \label{fig:uppper_limit_Pmax_G}
\end{figure}

We do not know the exact values of $o_n$ because distribution of the expected values of gas particles’ positions in relation to axis $S{-}D$ generally is unknown. Values stemming from the properties of the given physical setup should be used in the relevant calculations. For further considerations we will assume a random, uniform distribution of particles.
The graph in the FIG.~(\ref{fig:uppper_limit_Pmax_G}) shows $P_{G\gamma}^{max} (\bar{t})$. One can see that the spreading of gas particles progressing over mean free time $\bar{t}$ causes asymptotic driving to zero of the probability that photon $\gamma$ scattering will affect a finite size detector measurement. It is equivalent to the fact that \textit{the transmittance of such gas increases with extended mean free time}. Note the plateau at the beginning of the graph. It may be interpreted as the range of the applicability of the classical (not non-local) transmittance models e.g., Beer-Lambert law.

An interesting conclusion is that the \textit{measured transmittance of the smeared gas is dependent on the detector’s area} (see notes following Eq.~(\ref{LimesP})). Let us emphasize that the detector’s size not only has the obvious impact on the number of detected photons. Its size also affects the intrinsic optical properties of the smeared gas! It may be interpreted as a certain type of observer’s impact on the properties of the quantum system.

The product in Eq.~(\ref{TRUpperLimit}) is very complex computationally due to $N$ being a very high value, equal to the number of particles in the cloud. There is a simplified version, allowing for conducting calculations with any accuracy assumed, proposed in the Appendix.

Electromagnetic collisions $A_n{-}A_m$, i.e., of individual gas particles with one another, may be considered as a compounding of collision with an intermediate, at least one, photon: $A_n{-}\gamma{-}A_m$. Therefore, the above analysis are applicable.

\section*{Derivation of the Beer-Lambert law from the transmittance of the smeared gas}

Eq.~(\ref{TREquation}) describing the transmittance of the smeared gas should not contradict the ``classic'' transmittance equations e.g, the Beer-Lambert law \cite{A.D.McNaught1997}, within the specific range of applicability, i.e., for the non-smeared gas. Non-smeared gas is a gas where a particle’s mean free time is so short that there is not enough time for significant spontaneous spreading of a particle’s wave function to occur. Furthermore, the Beer-Lambert law assumes the use of a macroscopic detector. Both conditions are met when the standard deviation of the spreading $stdev_{A_n} (\bar{t})$ is significantly smaller than the detector’s diameter $\sqrt{A_D}$:
\begin{equation}
stdev_{A_n} (\bar{t}) \ll \sqrt{A_D}~, \label{BLDerivLargeDetecor}
\end{equation}
and $stdev_{A_n} (\bar{t})$ is significantly smaller than length $l$ of the sample (light ray length):
\begin{equation}
stdev_{A_n} (\bar{t}) \ll l~. \label{BLDerivLargeLength}
\end{equation}

Let’s derive the classic Beer-Lambert law equation from Eq.~(\ref{TREquation}) for a gas that is not subject to smearing i.e., fulfills Eq.~(\ref{BLDerivLargeDetecor}) and Eq.~(\ref{BLDerivLargeLength}). The transmittance of the smeared gas is impacted by all $N$ particles of the gas cloud, and not just the particles whose expected values of position are located within certain volume $V$ (of the detectability tunnel), in which the measurement of the transmittance is performed. Let us split $N$ particles assuming Eq.~(\ref{TREquation}) into a group located within volume $V$ (subject to the measurement of the transmittance) and the other group located outside this volume. Let us designate, as $N_V$, the number of particles in the group located within volume $V$. Then we can rewrite equation Eq.~(\ref{TREquation}) in the following form:
\begin{equation}
TR(\bar{t}) = \prod_{n=1}^{N_V} \big( 1-P_{A_n \gamma} (\bar{t}, o_n) \big) \prod_{n=N_V+1}^N \big(1-P_{A_n \gamma} (\bar{t}, o_n) \big)~.
\end{equation}

The second product is responsible for the particles whose expected values of position are located outside volume $V$: $o_n \gg r_T$. Due to the fact that the probability of their occurrence within volume $V$ is equal zero, we put $P_{A_n \gamma} (\bar{t}, o_n)=0$, and thus the second product is equal to $\sim 1$. We may then write
\begin{equation}
TR(\bar{t}) = \prod_{n=1}^{N_V} \big( 1-P_{A \gamma} (\bar{t}, o_n) \big)~. \label{TR_BL_InsideV}
\end{equation}

Let us then divide length $l$, i.e., between the source of light (beginning of the sample) and the detector, into slices in such a way that there is exactly one particle from the first group in each slice. The number $N_V$ of particles is finite so the number of slices will also be finite. Due to the fact that the particles are responsible for a potential photon absorption independent of one another and using Eq.~(\ref{BLDerivLargeLength}) we may allow ourselves to virtually shift them to the right or to the left (along $l$), so that the slices could have a uniform thickness $d_z$ each. This way we have $N_V=l/d_z$ slices.

Assuming Eq.~(\ref{BLDerivLargeDetecor}), considering geometry $G()$ of the setup (where the incident photons are almost parallel and source is of the cross section of detector) and using Eq.~(\ref{P_Ag}) we will determine that the probability of absorption \textit{detection} for each single particle is:
\begin{equation}
P_{A_n\gamma}(\bar{t}, o_n) = \frac{\sigma_{A_n\gamma}}{A_D}~,
\end{equation}
given some constant light wavelength. Simply the single particle covers detector with its (wavelength dependent) cross section.

Within volume $V=lA_D$ there are $N_V=l/d_z$ particles in total. So the number of particles $n_V$ within a unit of volume reaches:
\begin{equation}
n_V=\frac{N_V}{V}=\frac{l/d_z}{lA_D} =\frac{1}{d_z A_D}~,
\end{equation}
which we will insert into the previous equation and obtain:
\begin{equation}
P_{A_n\gamma}(\bar{t}, o_n) = \sigma_{A_n\gamma} d_z n_V~.
\end{equation}

Let us insert it into Eq.~(\ref{TR_BL_InsideV}) to get $\bar{t}$ independent transmittance $TR$:
\begin{equation}
TR = \prod_{n=1}^{N_V} \left ( 1-\sigma_{A_n\gamma} d_z n_V \right )~.
\end{equation}

Due to the fact that the expression in the product is dependent neither on $n$, nor on $\bar{t}$, and by inserting $N_V=l/d_z$, we may write:
\begin{equation}
TR = \left ( 1-\sigma_{A_n\gamma} d_z n_V \right )^{l/d_z}~.
\end{equation}

Applying the last equation to the definition of absorbance
\begin{equation}
ABS=-log_{10}(TR)=-ln(TR)/ln(10)
\end{equation}
and simplifying we obtain
\begin{equation}
ABS = - \frac{l~ln( 1-\sigma_{A_n\gamma} d_z n_V)}{d_z ln(10)}~.
\end{equation}

Expanding the logarithm into a series in relation to $d_z$ and disregarding higher order terms, we obtain the form of the classic Beer-Lambert equation we have been seeking:
\begin{equation}
ABS = - \frac{l~ ( -\sigma_{A_n\gamma} d_z n_V)}{d_z ln(10)} = \frac{l \sigma_{A_n\gamma} n_V}{ln(10)}~,
\end{equation}
which ends the derivation of the Beer-Lambert law from the transmittance of the smeared gas equation.

\section*{Possible occurrence of smeared gas}
We have shown above that matter in the form of smeared gas may take a form that is ``invisible'' to a \textit{single, small} observer. Let us then analyze how such a form of matter will absorb energy coming either from a point source or from the (e.g. cosmic) background radiation \cite{Holberg1986} \cite{MathisJ.S.MezgerP.G.Panagia1983}. This will allow us to determine whether the occurrence of stable smeared gas, concerning extended mean free time, is possible at all. The following preconditions are required to be met for its existence:
\begin{itemize}
\item low density of particles $A_n$, i.e., very rare internal interactions,
\item low intensity of external sources of other particles (e.g. photons) potentially colliding with particles $A_n$, i.e., weak external sampling,
\item appropriate time for the spreading of the wave functions of $A_n$.
\end{itemize}
These conditions can be met in deep space vacuum, at appropriate distances from radiation sources.

Let us try to determine what conditions must be met for the smearing of gas to occur and for such smearing to be sustained, or even expand, over time. Let us disregard the condition of a rare occurrence of collisions $A_n{-}A_m$ in the smeared gas. Let us assume that they have a significantly smaller impact on the reducing extended mean free time than collisions $A_n{-}\gamma$. We will justify this statement by asserting that the stipulated density of such gas is very low and the time between potential collisions, even of non-spread particles $A_n$, may be quite long in deep space \cite{Ferriere2001}. For example, in the Solar System vicinity \cite{Oliveetal.ParticleDataGroup2014} the density of all matter (baryonic $\&$ dark) is assumed to be ${\sim}4.1{\times}10^5$ proton masses per cubic meter. Assuming that i) even all this gas is baryonic and ii) is primarily made up of light components ${^1}\textrm{H}$, ${^2}\textrm{H}$, $\textrm{H}_{3}^{+}$ (${\sim}90\%$) and ${^3}\textrm{He}$, ${^4}\textrm{He}$ (${\sim}10\%$), we may roughly estimate the number of gas elements as ${\sim}2{\times}10^5$ per square meter. With the area of the total cross-section of the above mentioned particles in the region of ${\sim}10^{-11} [\textrm{m}^2]$ and quite low, non-relativistic speeds, the mean time between collisions $A_n{-}A_m$, involving even non-spread particles, is very long. With the classical ideal gas equations the mean free path is ${\sim}10^{16} \textrm{[m]}$. Taking a non-relativistic particle’s speed (e.g., ${\sim}0.01c$), the mean free time turns to be in the order of hundred years.

\subsection*{Occurrence of smeared gas at a given distance from radiation source}

In order to model photons spreading out from the source we will use the model of waves of probability of the occurrence of collisions spreading out from the source. Identical results are obtained for the model of photons as corpuscle particles fanning out from the source. Therefore, we will skip the derivation of formulas in the corpuscle model.

\subsubsection*{Monochromatic source}

We will cover the estimates of the preconditions that must be met for collisions $A{-}\gamma$ not to occur too frequently, allowing for the particles $A$ sustainable smearing. We are conducting the discussion in the cosmic scale, so we will assume that the stars are the sources of photons $\gamma$ in the Universe. To not lose the generality of this discussion we may assume that they are point sources $S$ with fixed brightness: within a unit of time the source $S$ emits constant energy $E_S$. It is spreading out in the form of electromagnetic waves, providing a certain probability of a collision occurring between photons and the particles encountered. The waves are spreading out spherically from source $S$. Let us also assume time independent spatial and spectral distribution of the radiation. Further analysis will be related to a freely selected, fixed photon wavelength, i.e., a narrow band of the source spectrum. We will designate the energy emitted by source $S$ in band $\lambda$ as $E_{S\lambda}$.

Knowing energy $E_{\lambda}$ of a photon of a specific wavelength one may calculate number $N_{\lambda}$ of photons of a specific wavelength, emitted by the source $S$ within a unit of time:
\begin{equation}
N_{\lambda}=\frac{E_{S\lambda}}{E_{\lambda}} \label{N_lambda}
\end{equation}
denominated in the reciprocal of the unit of time. We assumed it’s constant over time.

Let us again assume the Cartesian coordinates with axis $Z$, parallel to axis source $S$ – the geometric center of the area of the occurrence of particle $A$, and the beginning in the geometric center of area $A$. Let us now calculate the probability of collision $A{-}\gamma$. The particle $A$ probability density standard deviation is $stdev_A (\bar{t})$ as usual, its position expected value is located at distance $d$ from the source. Photon $\gamma$ sent from the source creates the wave of the probability of the occurrence of a collision with a spherical front. We will assume its point thickness in the outbound direction. Due to the normalization to one of the probability of the occurrence of photon $\gamma$ in the entire space, one can see that the probability of any photon collision on the unit area of the dome at distance $d$ from the source must take into account the factor of the probability of the occurrence of photon $P_{\gamma}(d)$ there, that reaches:
\begin{equation}
P_{\gamma}(d)=\frac{1}{4\pi d^2}
\end{equation}
denominated in the reciprocal of the unit of area.

Let us also assume that distance $d$ is significantly larger than standard deviation $stdev_A (\bar{t})$ of the particle $A$ spreading for every interesting $\bar{t}$: $d \gg stdev_A (\bar{t})$. We may then, for the simplicity of the formulas, disregard the change of the probability during the propagation in outbound direction $Z$ and assume that the upcoming photon waves are flat along axis $Z$. Then, with a sufficient approximation, one may define dimensionless probability $P_{AS} (\bar{t},d)$ of the occurrence of collision $A{-}\gamma$ (with photons $\gamma$ from $S$) at distance $d$ from source $S$:
\begin{equation}
P_{AS} (\bar{t},d) = \sigma_{A\gamma} \int_{-\infty}^{\infty} |\Psi_A(\mathbf{r},t)|^2 P_{\gamma}(d)dr = \frac{\sigma_{A\gamma}}{4\pi d^2}~.
\end{equation}

Index $AS$ denotes the probability of a collision of particle $A$ with photon coming from source $S$ within the entire potential space of the occurrence of particle $A$, in contrast to the previously calculated probabilities $P_{A\gamma}$ of the occurrence of collision $A{-}\gamma$ in the detectability tunnel. In accordance with intuition, this probability is not dependent on (extended mean free time) $\bar{t}$, but solely on distance $d$ between the geometric center of the area of particle $A$ and source $S$ and on the total cross-section of collision $A{-}\gamma$.

Now we may calculate the expected value of period $\bar{t}_{AS}$ between successive collisions $A{-}\gamma$ in the function of distance $d$ from source $S$. With constant number $N_{\lambda}$ of photons emitted by the source within a unit of time Eq.~(\ref{N_lambda}), the expected number of collisions is $N_{\lambda} P_{AS} (\bar{t},d)$, so $\bar{t}_{AS}$ will reach:
\begin{equation}
\bar{t}_{AS} (d) = \frac{1}{N_{\lambda} P_{AS} (\bar{t},d)} = \frac{4\pi E_{\lambda}}{\sigma_{A\gamma} E_{S\lambda}} d^2 = \alpha_{AS}d^2~. \label{t_AS}
\end{equation}

The quotient in the above formula is dependent solely on source $S$ (energy and spectrum) and type of particle $A$. We designated it in brief as $\alpha_{AS}$, and it is expressed in units of time per area, in the SI system $[\textrm{s}~\textrm{m}^{-2}]$.

We note that the expected value of period $\bar{t}_{AS}$ is constant over time and it is dependent solely on distance $d$ from the given source. We also note that the average time between successive collisions increases in proportion to the squared distance from source $S$.

Let us note that such areas of space may exist, at large distances from light sources. Time $\bar{t}_{AS}$ may be greater than the age of the Universe. The age of the Universe set the upper limit of the particles $A$ wave functions spreading time there. Estimate calculations indicate that such distances will be measured in light years.

\subsubsection*{Wide spectrum source}

The calculations for a wide spectrum source, for example a star, should be performed using the above scheme, dividing the entire spectrum of the source with required resolution. Source $S$ with energy $E_S$ emitted over a broad spectrum should be treated as a sum of energy $E_{S\lambda}$ of multiple sources with narrow bands: $E_{S\lambda}{=}\sum_{\lambda} E_{S\lambda}$. Then, the formula~(\ref{t_AS}) can be provided for various ranges of wavelengths $\lambda$, energies $E_{\lambda}$ and the probabilities of the occurrence of collision $P_{AS} (\bar{t},d,E_{\lambda})$ related thereto:
%\begin{widetext}
\begin{equation}
\bar{t}_{AS} (d) = 
\frac{1}{\sum_{\lambda}{\left[ N_{\lambda} P_{AS} (\bar{t},d,E_{\lambda}) \right]}} = 
\frac{4\pi}{\sum_{\lambda}{\left[ N_{\lambda} \sigma_{A\gamma}(E_{\lambda}) \right]}} d^2 = 
\frac{4\pi}{\sum_{\lambda}{\left[ \sigma_{A\gamma}(E_{\lambda}) E_{S\lambda}/ E_{\lambda} \right]}}  d^2 = 
\alpha_{AS}d^2~,
\end{equation}
%\end{widetext}
where the summation is taking place along the width of the spectrum in narrow bands around wavelength $\lambda$ each, $N_{\lambda}$ is the number of photons emitted within a unit of time in the band around wavelength $\lambda$, $E_{S\lambda}$ is the energy of the source within band $\lambda$ emitted in a unit of time and $\sigma_{A\gamma}(E_{\lambda})$ is the constant total cross-section for collision $A{-}\gamma$ with a photon with energy $E_{\lambda}$. We have used $\alpha_{AS}$ as a designation of the constant proportionality coefficient again. The narrower the bands the more accurate the calculation result will be.

\subsubsection*{Background radiation}

Apart from the source radiation, also the background radiation \cite{Holberg1986} \cite{MathisJ.S.MezgerP.G.Panagia1983} will usually have to be taken into account particularly while examining the spreading inside galaxies. Let us assume that radiance $I_{IGL} (\lambda)$ of such radiation is dependent on the wavelength but it is uniform in the entire area of the occurrence of particle $A$. For the radiance expressed in $[\textrm{W}~\textrm{m}^{-2}~ \textrm{sr}^{-1}]$ when summing the radiation from the entire background we must take into account $4\pi$ factor. The average number of reactions $N_{AI}$ of particle $A$ (within a unit of time), caused by background radiation $I_{IGL} (\lambda)$, is then:
\begin{equation}
N_{AI}=4\pi \sum_{\lambda}{\left[ \sigma_{A\gamma}(E_{\lambda})  I_{IGL} (\lambda)/E_{\lambda} \right]}~,
\end{equation}
which should be taken into account in above equations for $\bar{t}_{AS}$:
\begin{equation}
\bar{t}_{AS} (d) = \left(\frac{1}{\alpha_{AS}d^2} + N_{AI} \right)^{-1}~.
\end{equation}

It is easy to state now what the expected spreading of particle $A$ in the function of distance $d$ from source $S$ for a single wavelength $\lambda$ will be. We will provide it as the expected value of standard deviation $\overline{stdev_A} (d)$ in accordance with Eq.~(\ref{Stdev}):
%\begin{widetext}
\begin{equation}
\overline{stdev_A} (d) = 
\sqrt{\frac12 \left( \Delta^2+ \left( \frac{\hbar \bar{t}_{AS}(d)}{m\Delta} \right)^2 \right)} = 
\sqrt{\frac12 \left( \Delta^2+ \left( \frac{\hbar}{m\Delta} \right)^2 \left( \frac{1}{\alpha_{AS}d^2} + N_{AI} \right)^{-2} \right)}~.
\end{equation}
%\end{widetext}

\section*{Falsification}

At least several methods of carrying out an experiment falsifying the proposed model may be proposed.
\begin{itemize}
\item Transmittance of the laboratory created smeared gas can be measured with Tunable Diode Laser Absorption Spectroscopy (TDLAS) techniques. A sufficiently small sensor active area should be considered. Measurements of the transmittance of the same gas medium using detectors with various active areas could be performed for falsifying the transmittance vs. detector area relationship \cite{Ratajczak2020}.
\item Changes of the transmittance of gas at various distances from stars may be examined and compared to the above predictions. In particular, the geocorona Sun radiation dependence phenomena could be scrutinized more thoroughly.
\item If it turns out that distance from the Sun at which detectable smearing of the interstellar gas begins to occur is sufficiently short and thus located within the reach of direct exploration – a space probe equipped with a device locally sampling the transmittance of the gas could be sent.
\item The above equations could be used for developing cosmological and astrophysical models (forming of the matter systems: galaxies, stars or planetary systems). The simulations and the results obtained could be verified against observations.
\end{itemize}

One of the above proposed experiments was conducted. Namely, a TDLAS experiment that compares transmittances of ultra thin gas measured in parallel with a pair of detectors with different diameters \cite{Ratajczak2020}. Qualitatively it is in agreement ($>5\sigma$) with predictions of this paper.

\section*{Summary}

The smeared gas model presented herein may be related to several physical phenomena. 
\begin{itemize}
\item \textit{Geocorona}
- The observed increase of the geocorona \cite{Baliukin2019} absorbance from the Sun’s side and the correlation between Sun storms and the geocorona absorbance \cite{Kuwabara2017} \cite{Zoennchen2017} may, potentially, be explained using the presented model. As the geocorona exposed to stronger light will be subjected to more frequent collisions with the Sun’s photons, the extended mean free time will be shorter and therefore the geocorona will be more collision prone, which will be visible as a local transmittance drop which \textit{seems to be} local increase of density.
\item \textit{Dark matter} - may consist, among others, of baryonic smeared gas. Particles of such gas retain their mass, while at the same time they become transparent for electromagnetic waves and for one another. Note that it doesn't exclude other, non-baryonic dark matter components.
\item \textit{Possible matter and antimatter coexistence}
- Smeared intergalactic gas constitutes a perfect insulator between matter and antimatter. As the above described principles of the decreasing of the probability of collisions are also applicable to the particle-antiparticle pairs’ annihilation collisions. This way matter may blend with antimatter in the wide interstellar spaces, not annihilating each another.
\item \textit{Kuiper’s Cliff}
- Planetary systems might have been formed around active stars not only due to their gravitational pull, but also due to the fact that a star’s electromagnetic radiation caused sufficient reduction of the mean spreading of particles around the star and, as a consequence, it allowed for the occurrence of reactions leading to the merging of the protoplanetary matter at all. The so-called Kuiper’s cliff \cite{Larsen2007}, i.e. a ``sudden'' drop in the number of celestial bodies on the outer border of the Solar system, is observed. 
\end{itemize}
It's not claimed that smeared gas is responsible for enlisted phenomena. This list is brought up just to show some \textit{possible} applications and implications of the proposed model. No doubt, a careful research, which is out of scope of this paper, is required.

This model requires the full non-locality of wave functions. Experimental results following its predictions may be an important voice in the discussion on the interpretations of Quantum Mechanics. In particular the model contradicts interpretations postulating localized particles e.g., the pilot wave interpretations.

\subsection*{Conclusions}
Based on the quantum mechanical equations for non-relativistic particles, a theory envisaging a significant reduction of the absorption of very diluted gases (or an effective total cross-section drop) due to particle wave functions spreading has been proposed. This theory can be falsified through the proposed experiments. It may refer to a number of physical phenomena observed. Formulas for calculating the electromagnetic transmittance have been proposed. The classical Beer-Lambert law was derived from proposed formulas and the range of the applicability of this law has been outlined. It was shown that such a form of matter may be sustainable in deep space conditions.
Considerations in the paper can be generalized with respect to other types of collisions of various physical objects represented by wave functions, even for macroscopic objects. They are not restricted to a photon scattering only.

There are reports of experimental results consistent with the predictions of this model \cite{Ratajczak2020}.

\section*{Supporting information}
\paragraph*{S1 Appendix.}
\label{S1_Appendix}
{\bf Simplifying the formula of the smeared gas transmittance.}

The product appearing in the Eq.~(\ref{TREquation}) defining the transmittance of the smeared gas
\begin{equation}
TR(\bar{t}) = \prod_{n=1}^N \big( 1-P_{A_n \gamma} (\bar{t}, o_n) \big)
\end{equation}
has a computational complexity of $O(N)$, where, let us recall, $N$ is the number of particles in the cloud and - describing the situation in the orthodox terms – even in the entire Universe. It may be extremely large number. Probability $P_{A_n \gamma} (\bar{t}, o_n)$, calculated for each particle separately, requires a numerical integration or, even with the use of the approximations proposed in this paper, computing at least two $erf()$ functions and performing a number of additional arithmetic operations, see Eq.~(\ref{PmaxAnalytic}).

Therefore, we will propose a simplified version of this formula, allowing for conducting the calculations with any assumed accuracy. Although this proposal assumes an isotropic distribution of the particles in the cloud, it could be easily adapted to other distributions.

First of all, we will assume that the particles positioned too far to have a real impact on the transmittance measured can be disregarded in the calculations. Let us set certain radius $r$, running from axis $S{-}D$ source-detector. We will usually be performing the calculations for a cloud with a relatively concentrated distribution of standard deviations $stdev_{A_n}(\bar{t})$. Therefore, let us disregard the impact of the particles whose expected value of position ${\langle \hat{X} \rangle}_A$ is located at a distance greater than $r$. Parameter $r$ can be chosen arbitrarily, but it should meet condition $stdev_{A_n}(\bar{t}) \ll r$.

Let us assume that the source and the detector are sufficiently distant from the cloud to disregard the impact of the particles from behind the source and from behind the detector. Therefore, we obtain a cylinder with a radius of the base $r$ stretching from the source to the detector. The cylinder’s axis is detectability tunnel $T$. Let $N_r$ denote the number of particles that we will take into account, i.e. whose expected value of position ${\langle \hat{X} \rangle}_A$ is located inside this cylinder.

Let us note the symmetry of the impact of the particles on the transmittance in the tunnel. The particles positioned at an identical distance from the tunnel axis have an identical impact on the transmittance. The cylinder can be divided into disjunctive coaxial hollow cylinders (shells) so that all of them could fill up the cylinder. Thus we will obtain $K$ shells. Each of shells $k$ will have a certain thickness $\epsilon_k$ and distance from axis $o_k$. Any method of selecting shells is allowed, their thickness may, as a principle, vary. In order to increase the accuracy of the calculations, due to the fact that particles positioned closer to the axis have a greater impact on the value of the transmittance, it is advisable that the shells positioned closer to the axis of the cylinder should be thinner. To simplify, we will assume a fixed shell thickness $\epsilon{=}\epsilon_k$.

We will assign to the individual shells the particles whose expected value of position ${\langle \hat{X} \rangle}_A$ is located within the shell volume. The expected values of position ${\langle \hat{X} \rangle}_A$ of the particles assigned to shell $k$ are located, with the accuracy of $\pm \epsilon /2$, at distance $o_k$ from the axis:
\begin{equation}
o_k=\frac{r}{K} \left( k-\frac12 \right)~.
\end{equation}

Note that, the formula for distances $o_k$ will be different for other than uniform division of the shells.

Each shell has $N_k$ particles assigned thereto. One may calculate this number using the distribution of the expected values of position of particles ${\langle \hat{X} \rangle}_A$. This distribution is dependent, in an obvious manner, on the configuration of a specific physical setup. For example, for a uniform distribution within a volume limited by $r$ the number of particles $N_k$ in each shell will reach:
\begin{equation}
N_k = \frac{2N_r o_k \epsilon_k}{r^2} = \frac{N_r}{K^2}(2k-1)~.
\end{equation}

Using the independence of absorption occurrences again the formula for the transmittance of the smeared gas Eq.~(\ref{PmaxAnalytic}) may be approximated to:
\begin{equation}
TR(\bar{t}) \cong \prod_{k=1}^K \big( 1 - P_{A_n \gamma} (\bar{t}, o_k) \big)^{N_k}~.
\end{equation}

This formula has a computational complexity of $O(K)$, but since the number of shells $K$ is significantly lower than the number of particles $N$ ($K \ll N$), we are gaining an enormous acceleration of the calculations.
The accuracy of the calculations is dependent on the assumed value of $r$ and the number of shells $K$, as well as their thickness $\epsilon_k$. The accuracy can be improved by increasing $r$ and $K$, as well as by selecting appropriate $\epsilon_k$.

\nolinenumbers

%\bibliography{library.bib}

\end{document}